\documentstyle[12pt,rotate]{article}
\newcommand{\be}{\begin{equation}}
\newcommand{\ee}{\end{equation}}
\newcommand{\bea}{\begin{eqnarray}}
\newcommand{\eea}{\end{eqnarray}}
\newcommand{\bd}{\begin{displaymath}}
\newcommand{\ed}{\end{displaymath}}

\begin{document}
\bibliographystyle{physics}
\renewcommand{\thefootnote}{\fnsymbol{footnote}}

\author{
 Hongying Jin
\thanks{Email: jhy@hptc5.ihep.ac.cn, 
}\\
{\small\sl  Institute of High Energy Physics, Academia Sinica,
P.O.Box 918(4), Beijing 100039, China\thanks{Mailing address} }\\
}
\date{}
\title{
{\large\sf
\rightline{BIHEP-Th/98-009}
}
\vspace{3cm}
\bigskip
\bigskip
{\LARGE\sf Final State Interaction in $B\rightarrow K\pi$ Decays }
 } 
\maketitle
\begin{abstract}
\noindent
 We discuss the  rescattering effects in decays $B\rightarrow \pi+K$.  
The picture we take is very simple: first, B decay into
$K*\rho$, then $K*$ and $\rho$ go to kaon and pion by exchanging a pion.
We find a up to ten percent CP violation asymmetry rate. We also
discuss its correction to the constraint of angle $\gamma$ proposed 
recently. 
\end{abstract}
\vspace{1.5cm}
{\bf PACS  numbers 13.25.Hw 12.28.Lg } 

\newpage

\noindent{\bf\large  Introduction}\\

B meson's weak decay plays an important role in Standard  Model. It
provides a tool to testing  unitarity of  CKM matrix and the
possibility of  the first CP violation evidence out of the kaon 
system. However, the relation between the experimental observables 
and theoretical parameters sometimes is not clear, for instance, in the 
$B$ rare decay, the tree level amplitudes and one loop penguin amplitudes 
both give competing contributions, this situation makes it  difficult to
extract  the  angle $\gamma$ 
($arg[-\frac{V_{ud}V^*_{ub}}{V_{cd}V^*_{cb}}]$) of the unitary triangle 
from $B$ decays.

Recently CLEO collaboration  has presented 
the branching ratios\cite{CLEO} 
\begin{eqnarray}
\displaystyle{\frac{1}{2}
[Br(B^0\rightarrow\pi^-K^+)+Br({\bar B}^0\rightarrow\pi^+K^-)]
=(1.5^{+0.5}_{-0.4}\pm 0.1\pm 0.1)\times 10^{-5}}\\        
\displaystyle{\frac{1}{2}
[Br(B^-\rightarrow \pi^-K^0)+Br(B^+\rightarrow \pi^+{\bar K}^0)]
=(2.3^{+0.1}_{-1.0}\pm 0.3\pm 0.2)\times
10^{-5}}.
\label{data}
\end{eqnarray}        
It  attracted much interest. On the one hand, (\ref{data}) may 
provide a  constraint of the angle 
$\gamma$. As pointed by Fleischer and Mannel\cite{flei},   
using the branching ratios of these four $B\rightarrow \pi K$ decay modes, 
it is possible to derive
a bound on the angle $\gamma$ of the unitarity triangle which, under 
certain circumstance, is free of hadronic uncertainties. 
On the other hand, it is believed that the direct CP violation asymmetry 
is small in $B^\pm\rightarrow K\pi^\pm$, so (\ref{data}) can be used 
to search new physics. However, these two arguments are based on
perturbative QCD, the long distance strong interaction such as 
final state interaction may not be negligible.   
As discussed in \cite{f1}, after taking account of  
$K\pi\rightarrow K\pi$ rescattering ,
the recent observes (\ref{data}) do not lead to a significant bound 
on the angle $\gamma$; Besides, authors claim that   
a sizable CP violation asymmetry rate is possible in 
$B^\pm\rightarrow \pi^\pm K$.  More rescattering channels such as 
$\eta K$,$K^*\pi\rightarrow \pi K$ have been 
discussed in \cite{f2,f3,f4}, the authors also found a $10-20\%$
correction to the up bound of $\gamma$ and $O(10\%)$ CP violation
asymmetry rate.
    
The final state interaction in $B\rightarrow K\pi$ can be through many
channels, for instance, the rescattering channels  
$B\rightarrow VV\rightarrow K\pi$  have not been discussed yet.  
In this paper, 
we will discuss the  channel $B\rightarrow\rho K^*\rightarrow K\pi$, the 
reason is that the branching ratio of  $B\rightarrow \rho K^*$ may be  
larger than $B\rightarrow K\pi$, rescattering effect may not be small. 
Besides, in the rescattering process,
$B^-\rightarrow \rho^0 K^{*-}\rightarrow \pi^- {\bar K}^0$ 
the tree amplitude's contribution is non-zero, which can provide
a correction to the constraint of angle $\gamma$  and the 
direct CP asymmetry in the charge $B$ decay $B\rightarrow K^0\pi$.  

The final state interaction is a long distance strong interaction, 
how to dual with it in theory is not clear so far. A common method 
is Reggle pole model,which was used in \cite{f3,f4}. In our case, this 
model is not valid. However,  
we can use a rather similar estimate which has been used in
\cite{Li}. The picture is, first $B$
decays into $K^{*}\rho$, then $K*$ and $\rho$ go to $\pi K$ by exchanging
a pion instead of a vector meson such as $K^*$ and $\rho$, the coupling
of $\rho \pi\pi$ and $K^*K\pi$ can be determined by 
the decay  $\rho\rightarrow 2\pi$ and $K^*\rightarrow K\pi$. 
In order to give a numerical result, BSW model is used to calculate the 
form factor of $B\rightarrow K^{*}\rho$ and $B\rightarrow K\pi$. \\

\noindent{\bf\large Effects of Final state interaction}\\

In general, the amplitudes of the relevant 
$B\rightarrow K\pi$ decays may be represented as\cite{flei} 
\begin{eqnarray}
A({\bar B}^0\rightarrow \pi^-K^+)=A^0_P-A^0_Te^{-i\gamma}\\        
A({     B}^0\rightarrow \pi^+K^-)=A^{\bar 0}_P-A^{\bar 0}_Te^{i\gamma}\\        
A(B^-\rightarrow \pi^-{\bar K}^0)=A^-_P-A^-_Te^{-i\gamma}\\
A(B^+\rightarrow \pi^+K^0)=A^+_P-A^+_Te^{i\gamma}\\
\label{KP}        
\end{eqnarray}
where $A_P$ and $A_T$ are penguin and tree amplitudes respectively. 
QCD penguin keeps $SU(2)$ isospin symmetry, compared with it,
the electroweak penguin contribution is small. We can roughly think 
$A^0_P=A^{\bar 0}_P=A^-_P=A^+_P=A_P$; In the Standard Model, CP violation
rises from 
CKM matrix, therefore $A^0_T=A^{\bar 0}_T,~A^-_T=A^+_T$, 
$\delta_-=\delta_+$.  Defining $A^0_T/A_P=re^{i\delta_0}$,
$A^-_T/A_P=\epsilon e^{i\delta_-}$  one may get the rate
\begin{equation}
R=\displaystyle{\frac{\Gamma(B^0\rightarrow
\pi^-K^+)+\Gamma({\bar B}^0\rightarrow \pi^+K^-)}
{\Gamma(B^-\rightarrow \pi^-{\bar K}^0)+\Gamma(B^+\rightarrow \pi^+{
K}^0)}=
\frac{1-2rcos\gamma cos\delta_0 +r^2}
{1-2\epsilon\cos\gamma\cos\delta_- +\epsilon^2}}
\label{r}
\end{equation}
and direct CP asymmetry ${\cal A}^{dir}_{CP}\equiv{\cal A}^{dir}_{CP}
(B^-\rightarrow\pi^-{\bar B}^0)$ \cite{f2}  
\begin{equation}
{\cal
A}^{dir}_{CP}=\displaystyle{\frac{BR(B^+\rightarrow\pi^+K^0)
-BR(B^-\rightarrow\pi^-{\bar K}^0)}{BR(B^+\rightarrow\pi^+K^0)
+BR(B^-\rightarrow\pi^-{\bar K}^0)}=
\frac{2\epsilon\sin\gamma\sin\delta_-}{1-2\epsilon\cos\gamma\cos\delta_-
+\epsilon^2}}
\label{cp}
\end{equation}
If rescattering effects are not taken into account, tree amplitudes
have no contribution to charge $B$ decay, i.e. $A^-_P=0$.   
Then  one can minimizes R with respect to the parameter $r$ and obtain 
an inequality $sin^2\gamma\leq R$; Moreover, the direct CP asymmetry rate  
can be negligible.   

The rescattering process involves an intermediate on-shell $X$, 
such that $B\rightarrow X\rightarrow K\pi$. In this paper , we 
choose $X$ as $\rho K^*$. We denote  direct  amplitude 
of $B\rightarrow \pi K$ and rescattering amplitude of 
 $B\rightarrow \rho K^*\rightarrow\pi K$ as 
$A_{dir}$ and $A_{res}$ respectively. Similarly to (\ref{KP}), $A_{dir}$ 
and $A_{res}$  can be written as 
\begin{equation}
\begin{array}{l}
A^0_{dir}=A^0_P-A^0_Te^{-i\gamma},\\        
A^-_{dir}=A^-_P,\\        
A^0_{res}={\cal A}^0_P-{\cal A}^0_Te^{-i\gamma},\\
A^-_{res}={\cal A}^-_P-{\cal A}^-_Te^{-i\gamma}.
\end{array}
\end{equation}
Correspondingly, the parameters $r=|\frac{A^0_T+{\cal A}^0_T}{A^0_P+{\cal
A}^0_P}|$, $\epsilon=|\frac{{\cal A}^-_T}{A^-_P+{\cal A}^-_P}|$ and 
$\delta_-=arc[\frac{{\cal A}^-_T}{A^-_P+{\cal A}^-_P}]$.

 In the Standard Model, these decays are mediated by the $\Delta B=1$ 
Hamiltonian, which takes the form 
\begin{equation}
H_{eff}=\displaystyle{\frac{G_F}{\sqrt{2}}} 
[V_{ub}V_{us}^*(c_1O_1^u+c_2O_2^u)+V_{cb}V_{cs}^*(c_1O_1^c+c_2O_2^c)
-V_{tb}V_{ts}^*\sum_{i=3}^{10}c_iO_i]+h.c.,
\end{equation}

where 
\begin{equation}
\begin{array}{ll}
O_1^u=(\bar s b)_{V-A}(\bar u u)_{V-A}, &O_2^u=(\bar u b)_{V-A}(\bar s
u)_{V-A},\\
O_{3(5)}(\bar s b)_{V-A}\sum_{q'}({\bar q}' q')_{V-A(V+A)}, &
O_{4(6)}({\bar s}_\alpha b_\beta)_{V-A}\sum_{q'}({\bar q}'_\beta
q'_\alpha)_{V-A(V+A)},\\
O_{7(9)}=\displaystyle{\frac{3}{2}(\bar s b)_{V-A}\sum_{q'}e_{q'}
({\bar q}' q')_{V+A(V-A)}}, &
O_{8(10)}=
\displaystyle{\frac{3}{2}({\bar s}_\alpha b_\beta)_{V-A}\sum_{q'}e_{q'}
({\bar q}'_\beta q'_\alpha)_{V+A(V-A)}},
\end{array}
\end{equation}

with $O_1-O_2$ being current current operators, $O_3-O_6$ being the QCD
penguin operators and $O_7-O_{10}$ the electroweak penguin operators. 
To the next-to-leading order, the Wilson 
coefficients $c_i(\mu)$ depends on the renormalization scheme chosen,  
it is convenient to use the  effective
Wilson coefficients $c^{eff}_i$ instead, which is independent of the
renormalization scheme and defined by \cite{cheng}
\begin{equation}
c^{eff}_{i}=[1+\frac{\alpha_s(\mu)}{4\pi}{\bf m}^T_s(\mu)+
\frac{\alpha(\mu)}{4\pi}{\bf m}^T_e(\mu)]_{ij}c_j(\mu)
\end{equation} 
The numerical values of $c^{eff}_i$ have been given in \cite{cheng} in 
the 't Hooft Veltman (HV) scheme and naive dimension regularization scheme 
at $\mu=m_b(m_b)$,$\Lambda_{\bar{MS}}=225Mev$ and $m_t=170Mev$: 
\begin{equation}
\begin{array}{ll}
c^{eff}_1=1.149, &c^{eff}_2=-0.325,\\
c^{eff}_3=0.0211+i0.0045, &c^{eff}_4=-0.0450-i0.0136,\\
c^{eff}_5=0.0134+i0.0045, &c^{eff}_6=-0.0560-i0.0136,\\
c^{eff}_7=-(0.0276+i0.0369)\alpha, &c^{eff}_8=0.054\alpha,\\
c^{eff}_9=-(1.318+i0.0369)\alpha, &c^{eff}_{10}=0.263\alpha,
\end{array}
\end{equation}
Then, the factorization approximation can be applied to the hadronic 
matrix elements of the operator $O_i$ at the tree level. The direct 
decay amplitudes of $B\rightarrow K\pi$ and 
$B\rightarrow K^*\rho$ in factorization approximation can be written as
\begin{equation}
\begin{array}{lll} 
A_{dir}({\bar B}^0\rightarrow\pi^+K^-)
&=&\displaystyle{\frac{G_F}{\sqrt{2}}
\{V_{ub}V_{us}^*a_1+V_{cb}V_{cb}^*[a_3+\frac{3}{2}e_ua_9
\frac{2m_K^2}{(m_s+m_u)(m_b-m_u)}(2a_5+3e_ua_7)]\}M^2_{K^-\pi^+}}\\
A_{dir}(B^-\rightarrow\pi^-{\bar K}^0)
&=&\displaystyle{\frac{G_F}{\sqrt{2}}
V_{cb}V_{cb}^*[a_3+\frac{3}{2}e_da_9
\frac{2m_K^2}{(m_s+m_d)(m_b-m_d)}(2a_5+3e_da_7)]M^4_{{\bar K}^0\pi^-}}\\
A_{dir}({\bar B}^0\rightarrow {K^*}^-\rho^+)
&=&\displaystyle{\frac{G_F}{\sqrt{2}}
\{V_{ub}V_{us}^*a_1+V_{cb}V_{cb}^*[a_3+\frac{3}{2}e_ua_9]\}
M^2_{K^{*-}\rho^+}}\\
A_{dir}({\bar B}^0\rightarrow {\bar K}^{*0}\rho^0)
&=&\displaystyle{\frac{G_F}{\sqrt{2}}
\{V_{ub}V_{us}^*a_2M^1_{{\bar
K}^{*0}\rho^0}+V_{cb}V_{cb}^*[a_3+\frac{3}{2}e_da_9]M^4_{{\bar 
K}^{*0}\rho^0}\}}\\
A_{dir}( B^-\rightarrow {K^*}^-\rho^0)
&=&\displaystyle{\frac{G_F}{\sqrt{2}}
\{V_{ub}V_{us}^*(a_1M^2_{{\bar 
K}^{*-}\rho^0}+a_2M^1_{{\bar K}^{*-}\rho^0})
+V_{cb}V_{cb}^*[a_3+\frac{3}{2}e_ua_9]
M^2_{{\bar K}^{*-}\rho^0}\}}\\
A_{dir}( B^-\rightarrow {\bar K}^{*0}\rho^-)
&=&\displaystyle{\frac{G_F}{\sqrt{2}}
\{V_{cb}V_{cb}^*[a_3+\frac{3}{2}e_da_9]M^4_{{\bar 
K}^{*0}\rho^-}\}}\\
\end{array}
\end{equation}
  
where $a_i$ are defined as 
\begin{equation}
a_{2i-1}=\displaystyle{c^{eff}_{2i}+
\frac{c^{eff}_{2i-1}}{N_c},~~~~~~~~~~~~~
a_{2i}=c^{eff}_{2i-1}+\frac{c^{eff}_{2i}}{N_c}},
\label{amp}
\end{equation}

and  $M^i_{ab}$ are the matrix elements of the
operators $O_i$ inserted in the states of $ab$($K\pi$ pseudo-scalar,vector
mesons) and $B$. Obviously,   
the rate of electroweak penguin to $QCD$ penguin is only 
$A^{EW}_P/A^{QCD}_P\sim a_9/a_3\sim 1/30$, so we omit $A^{EW}_P$.
   
Using vacuum-saturation approximation, we can write, for instance,
 $M_1$  and $M_3$  as  
\begin{equation}
\begin{array}{lll}
M^2_{K^-\pi^+}&=&-i\langle K^-|{\bar s}u|0\rangle_{V-A}\langle \pi^+|{\bar
u}b|{\bar
B}^0\rangle_{V-A}=f_K(m_B^2-m_\pi^2)F_0(m_K^2)\\
M^2_{K^{*-}\rho^+}&=&-i\langle K^{*-}(e^*)|{\bar
s}u|0\rangle_{V-A}\langle\rho^+(\eta)|
{\bar u}b|{\bar B}^0\rangle_{V-A}\\
&=&
\displaystyle{
-f_{K^*}e^*_\mu\{
\frac{2}{m_B+m_{\rho}}i\epsilon^{\mu\nu\alpha\beta}\eta_\nu
p^B_\alpha p^{\rho}_\beta
V(m^2_{K^*})}\\
&&
\displaystyle{
+[\eta_\mu(m_B+m_\rho)A_1(m_{K^*}^2)
-\frac{\eta\cdot p^{K^*}}{m_B+m_\rho}(p^B+p^\rho)_\mu
A_2(m^2_{K^*})]\}}\\ 
\end{array}
\label{M}
\end{equation}

In (\ref{M}), we use the  definitions\cite{BSW}
\begin{equation}
\begin{array}{lll}
\langle X|j_\mu|B\rangle_{V-A}&=&\displaystyle{
(P^B+P^X-\frac{m_B^2-m_X^2}{q^2}q)_{\mu}F_1(q^2)
+\frac{m_B^2-m_X^2}{q^2}q_\mu F_0(q^2)},\\
\langle X^*(\eta)|j_\mu|B\rangle_{V-A}&=&\displaystyle{
\frac{2}{m_B+m_{X^*}}i\epsilon^{\mu\nu\alpha\beta}
\eta_\nu p^B_\alpha p^{X^*}_\beta V(q^2) 
+[\eta_\mu(m_B+m_{X^*})A_1(q^2)-\frac{\eta\cdot q}
{m_B+m_{X^*}}(p^B+p^{X^*})_\mu A_2(q^2)}\\
&&\displaystyle{
-\frac{\eta\cdot q}
{q^2}2m_{X^*} q_\mu A_3(q^2)]+\frac{\eta\cdot q}{q^2}2 m_{X^*}
q_\mu A_0(q^2)},
\end{array}
\label{form}
\end{equation} 
with $q=p^B-p^X (p^{X^*})$, where $X(X^*)$ is an arbitrary 
pseudo-scalar (vector) meson. The numerical values of the  form factors
in (\ref{form}) are estimated  by  using WBS model. 
  
Now let's discuss the amplitudes of $K^*\rho\rightarrow K\pi$.  
This soft process can be  described by 
the low energy  effective Lagrangian.  
As showed in fig.1, $K^*\rho$ can go to $K\pi$ by exchanging one pion. 
Only keeping the lowest dimension operators, we 
can write the interaction of $K^*K\pi$ and $\rho\pi\pi$ in term
of $SU(2)$ isospin symmetry as 

\begin{equation}
\begin{array}{lll}
L_{K^*K\pi}&=&ig_{K^*K\pi}{{\bf K}_\mu^*}^\dag
[{\bf \pi}\partial^\mu{\bf K}-\partial^\mu{\bf \pi}{\bf K}]+h.c.,\\
L_{\rho\pi\pi}&=&g_{\rho\pi\pi}
\epsilon_{ijk}{{\bf \rho}^i}^\mu\partial_\mu\pi^j\pi^k,
\label{lag}
\end{array}
\end{equation}
with
\begin{equation}
\begin{array}{l}
{\bf \pi}=\left (\begin{array}{cc}
\frac{\pi^0}{\sqrt{2}}&\pi^+\\
\pi^-&-\frac{\pi^0}{\sqrt{2}}
\end{array}\right ),~~~~~{\bf K}=\left (\begin{array}{c}
K^+\\K^0\end{array}\right ),~~~~~~{\bf K}^*_\mu=
\left (\begin{array}{c} {K^*}^+\\{K^*}^0\end{array}
\right )_\mu\\
\pi^\pm(\rho^\pm)=\displaystyle{\frac{1}{\sqrt{2}}
[\pi^1(\rho^1)\pm i\pi^2(\rho^2)],~~~~~~~~~~~~~~~~
\pi^0(\rho^0)=\pi^3(\rho^3)}.
\end{array}
\end{equation}

Using the interaction (\ref{lag}), we obtain the amplitudes of
$K^*\rho\rightarrow K\pi$ 
\begin{equation}
\begin{array}{lll}
\langle{\bar K}^0\pi^-|K^{*-}(e^*)\rho^0(\eta)\rangle=
g_{K^*K\pi}g_{\rho\pi\pi}\eta\cdot(p_{\pi^-}-q)e^*\cdot(p_{{\bar  
K}^0}+q)\displaystyle{\frac{i}
{q^2}}\\
\langle{\bar K}^0\pi^-|{\bar K}^0(e^*)\rho^-(\eta)\rangle=
\displaystyle{\frac{1}{\sqrt{2}}
g_{K^*K\pi}g_{\rho\pi\pi}\eta\cdot(p_{\pi^-}-q)e^*\cdot(p_{{\bar  
K}^0}+q)\frac{i}{q^2}}\\
\langle K^-\pi^+|K^{*-}(e^*)\rho^+(\eta)\rangle=
\displaystyle{\frac{1}{\sqrt{2}}
g_{K^*K\pi}g_{\rho\pi\pi}\eta\cdot(p_{\pi^+}-q)e^*\cdot(p_{  
K^-}+q)\frac{i}
{q^2}}\\
\langle K^-\pi^+|{\bar K}^{*0}(e^*)\rho^0(\eta)\rangle=
\displaystyle{-
g_{K^*K\pi}g_{\rho\pi\pi}\eta\cdot(p_{\pi^+}-q)e^*\cdot(p_{  
K^-}+q)\frac{i}
{q^2}},
\end{array}
\end{equation}
where $q=p_{K^*}-p_K$ is the momentum of the exchanged pion. 

Finally, the rescattering amplitude of $B\rightarrow K^*\rho\rightarrow 
K\pi$ are written as     
\begin{equation}
A^-_{res}=\displaystyle{
 \frac{1}{2}\sum_{K^*\rho}\int\frac{d^3{\bf p}^\rho}{(2\pi)^32E_\rho}
\frac{d^3{\bf p}^{K^*}}{(2\pi)^32E_{K^*}}
\langle {\bar K}^0\pi^-|K^{*}\rho\rangle 
\langle K^{*}\rho|B^-\rangle }, 
\label{res}
\end{equation} 
where $\sum_{K^*\rho}$ sums all possible intermediate states $K^*\rho$ 

The numerical calculation is carried out with the parameters and 
form factors \cite{BCW}
\begin{equation}
\begin{array}{lll}
m_b=4.7Gev~~~~~~~~~~~~,&m_s=0.15Gev~~~~~~~~~~~~~&m_{u,d}=0.01Gev,\\
f_{K^*}=0.2Gev^2~~~~~~~, &f_\rho=0.2Gev^2~~~~~~~~~~&N_c=3\\
F_0(q^2)=\frac{0.39}{1-q^2/m^2_{B^*}}  ~~~~~~~       , 
&A_1(q^2)=\frac{0.36}{1-q^2/m^2_{B}}
~~~~& A_2(q^2)=\frac{0.36}{1-q^2/m^2_{B}} \\
\end{array}
\end{equation}
The values of coupling constants $g_{K^*K\pi}$ and $g_{\rho\pi\pi}$ in 
(\ref{lag}) can be determined by the decay widths of $K^*\rightarrow K\pi$ 
and $\rho\rightarrow\pi\pi$ respectively,   
\begin{equation}
\begin{array}{lll}
\Gamma(K^*\rightarrow K\pi)&=&\displaystyle{
\frac{3g_{K^*K\pi}^2}{2}\frac{
[m_{K^*}^2-(m_K+m_\pi)^2]^{\frac{3}{2}}
[m_{K^*}^2-(m_K-m_\pi)^2]^{\frac{3}{2}}}
{48\pi m_{K^*}^5}=50Mev}\\
\Gamma(\rho\rightarrow K\pi)&=&\displaystyle{
g^2_{\rho\pi\pi}\frac{[m_{\rho}^2-4m^2_\pi]^{\frac{3}{2}}}
{48\pi m_\rho^2}=150Mev}
\end{array}
\label{coupl}
\end{equation}
However, (\ref{coupl}) cannot determine  their sign; This is not
sensitive in our discussion here, we just choose positive sign, it 
may be important when one wants to sum all rescattering effects.   
The momentum integral of $K^*\rho$ in (\ref{res}) cannot be performed 
in the total phase space. When $K^*$ and $K$ go back to
back, the momentum of 
the exchanged pion is very large,$|p_{ex}|=5Gev$. On the other hand, final 
state interaction is a long distance process , we describe it by using 
the low energy effective lagrangian (\ref{lag}) which is not valid at very
high energy. So we need a cut-off $p_c$ to distinguish soft and 
hard regions. We choose $p_c=2Gev$, since in a short distance process,
$K^*$ and $\rho$ should exchange at least two gluons, then they can go 
to $K$ and $\pi$, the typical momentum of each gluon is $\sim 1 Gev$.  We
do the  intergation in the region $|p_{ex}|<p_c$, when 
$|p_{ex}|>p_c$, $K^*$ and $\rho$ should exchange hard gluons instead,
this short distance process should be strongly
suppressed\cite{hard1-hard3} , however it
is  beyond  our discussion.   

The numerical results are showed in the table.1.   
\begin{center}
Table-1 Amplitudes of decays $B\rightarrow K\pi$
\begin{tabular}{|c|c|c|}
\hline 
& tree ($V_{ub}V_{us}^*$)&penguin($V_{cb}V_{cs}^*$)\\
\hline
$A_{dir}({\bar B}^0\rightarrow K^-\pi^+)$ &$-1.5\times 10^{-5}$
&$(10.1+i2.8)\times 10^{-7}$\\
\hline
$A_{dir}(B^-\rightarrow {\bar K}^0\pi^-)$
&$0$&$(10.1+i2.8)\times 10^{-7}$\\
\hline
$A_{res}({\bar B}^0\rightarrow K^-\pi^+)$
&$i3.3\times 10^{-6}$&$(0.73-i2.3)\times 10^{-5}$\\
\hline 
$A_{dir}(B^-\rightarrow {\bar K}^0\pi^-)$
&$i3.5\times 10^{-6}$&$(0.73-i2.3)\times 10^{-5}$\\
\hline 
\end{tabular}
\end{center}
Form table.1, we obtain a large  strong phase 
$\delta_-\sim\frac{\pi}{2}$, because compared to direct 
amplitudes, rescattering amplitudes almost have a $\frac{\pi}{2}$ phase
transition.   
If we take $|\frac{V_{cb}V_{cs}^*}{V_{cb}V_{cs}^*}|\sim 0.02$ \cite{f2}, 
$\epsilon\approx 0.06$. Substitute these values into (\ref{cp}) and 
keep in mind that $\gamma$ may be larger than $\pi/3$, we conclude 
there is a $10\%$ CP violation asymmetry rate, which is comparable 
with \cite{f3,f4}.  In our case, there is almost no
correction to (\ref{r}), the up bound of $\gamma$ is not changed.  

In  conclusions, we discuss the final state interaction in decays
$B\rightarrow K\pi$ 
via rescattering channel $K^*\rho\rightarrow K\pi$. By using BCW model 
and low energy effective lagrangian, we estimate that this channel gives 
a $10\%$ CP violation asymmetry rate, which is comparable with other 
rescattering channels. \\

\noindent{\bf\large Acknowledgement}

The author thanks Dr.Z.T.Wei and Dr. M.Z. Yang for useful discussions. 
He also thanks Dr. Frank Krueger for his pointing out a misprint.
This work is
supported in part by the National Natural Science
Foundation.

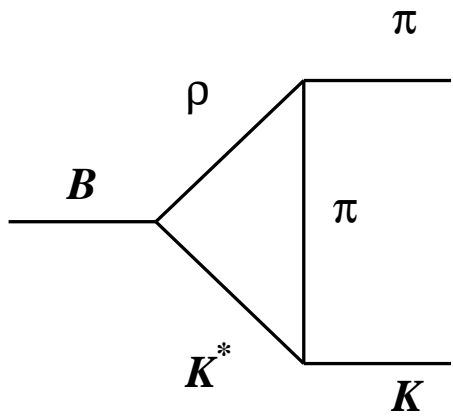
\end{document}